\documentclass[
reprint,
superscriptaddress,
showpacs,
preprintnumbers,
nofootinbib,
amsmath,
amssymb,
aps,
prd,
dvipsnames,
floatfix
]{revtex4-2}
\pdfoutput=1
\usepackage{xcolor}
\usepackage{slashed}
\usepackage{amsmath}
\usepackage{amsfonts}
\usepackage{amssymb}
\usepackage{mathrsfs}
\usepackage{slashed}
\usepackage{graphicx}
\usepackage[dvipsnames]{xcolor}
\usepackage{array}[=2016-10-06]
\usepackage{multirow,makecell}
\usepackage{listings}
\usepackage{upgreek}
\usepackage{bm}
\usepackage[normalem]{ulem}
\usepackage{booktabs}
\usepackage{xspace}
\usepackage{physics}
\usepackage{adjustbox}
\usepackage{chemformula}
\usepackage[inline]{enumitem}
\usepackage{fontawesome5} 
\usepackage[colorlinks=true,linkcolor=purple,citecolor=purple,allcolors=purple]{hyperref}
\usepackage[capitalise]{cleveref}
\usepackage{cancel}
\usepackage{dsfont}
\usepackage{comment}

\makeatletter
\AddToHook{cmd/appendix/before}{\def\cref@section@alias{appendix}}
\makeatother

\definecolor{orcidlogocol}{HTML}{A6CE39}
\usepackage{isotope}

\definecolor{brightpink}{rgb}{1.0, 0.0, 0.5}
\definecolor{emerald}{rgb}{0.31, 0.78, 0.47}
\definecolor{limegreen}{rgb}{0.2, 0.8, 0.2}
\definecolor{blue-violet}{rgb}{0.33, 0.17, 0.89}

\usepackage{makecell}
\usepackage{tabularx}
\newcolumntype{C}[1]{>{\centering\arraybackslash}m{#1}} 
\newcolumntype{P}[1]{>{\centering\arraybackslash}p{#1}} 
\newcolumntype{Y}[1]{>{\centering\arraybackslash\hsize=#1\hsize}X}

\definecolor{MH}{rgb}{0.0,0.6,0.9}

\begin{document}

\title{\Large Light on GeV-scale seesaw} 

\author{Enrico Bertuzzo}
\email{enrico.bertuzzo@unimore.it}
\affiliation{Dipartimento di Scienze Fisiche, Informatiche e Matematiche, Universit\`a degli Studi di Modena e Reggio Emilia, Via Campi 213/A, I-41125 Modena, Italy}
\affiliation{INFN, Sezione di Bologna, Via Irnerio 46, I-40126 Bologna, Italy}

\author{Michele Frigerio}
\email{frigerio@lpthe.jussieu.fr}
\affiliation{Laboratoire de Physique Th\'eorique et Hautes \'Energies (LPTHE), CNRS \& Sorbonne Universit\'e, 4 Place Jussieu, Paris, France}

\author{Salvador Urrea}
\email{salvador.urrea@ijclab.in2p3.fr}
\affiliation{IJCLab, P\^ole Th\'eorie (B\^at. 210), CNRS/IN2P3, 91405 Orsay, France}

\begin{abstract}
The next generation of beam-dump and accelerator-neutrino experiments will probe GeV-scale sterile neutrinos with
unprecedented sensitivity. However, the tiny active--sterile
mixing predicted by the minimal seesaw mechanism is expected to remain out of reach. Oscillation data require at least two sterile
neutrinos, $N_1$ and $N_2$, which may couple to the photon through a transition dipole. Such a coupling can radically change the picture,
both by strongly enhancing
sterile-neutrino production and by opening new decay channels. We identify charged-lepton pairs produced by an off-shell photon
as the most promising experimental signature.
We show that
the seesaw band can be probed by the T2K near detector for sterile-neutrino masses $m_N\lesssim 500$~MeV, and by SHiP for $m_N\lesssim 5$~GeV.
For a sufficiently large dipole,
$d\gtrsim (10~\mathrm{TeV})^{-1}$,
SHiP can selectively measure decay channels that require
nonzero active--sterile mixing. This would give rise to a distinctive
hierarchy of signatures, providing a
direct test of the seesaw mechanism.
\end{abstract}

\maketitle

\section{Introduction}

Gauge-singlet fermions, commonly
referred to as sterile neutrinos or heavy neutral leptons (HNLs), can exist at arbitrary mass scales, generically unrelated to the Standard Model (SM) one. 
Even though neutral, they can couple to the photon via a dipole interaction~\cite{Pauli:1941zz}, which could play a crucial role in their discovery, if they are sufficiently light to be directly produced in the laboratory. In this Letter, we highlight the potential of beam-dump experiments~\cite{Beacham:2019nyx,Bergsma:1986is,CooperSarkar:1985nh,SHiP:2018xqw} and the T2K near detector~\cite{T2K:2019hnl} to search for GeV-scale HNLs via their dipole coupling to photons. 
The dominant dipole-mediated decay produces a monophoton signal that is very challenging to distinguish from background~\cite{Ferrillo:2023hhg}. We will discuss how to circumvent this difficulty
at present and future detectors,
by looking for decays to charged-lepton pairs mediated by an off-shell photon, a much cleaner signature.

We focus on what is perhaps the best motivated and minimal model for HNLs, that is, two sterile neutrinos whose couplings to active neutrinos allow to reproduce the observed neutrino oscillation parameters, via the seesaw mechanism~\cite{Minkowski:1977sc,Yanagida:1979as,Gell-Mann:1979vob,Mohapatra:1979ia}.
We will show that, in the presence of a sterile-neutrino dipole, it becomes possible to probe an active-sterile mixing as small as predicted by the minimal seesaw, that is: 
\begin{equation}
U_{\nu N} \simeq \sqrt{\frac{m_\nu}{m_N}}~,
\label{mixI}
\end{equation}
where $m_\nu \sim 0.1$ eV is the observed scale of light neutrino masses and $m_N$ is the arbitrary mass scale of heavy neutrinos.
The mass range accessible to beam-dump experiments, $m_N\sim (10 \,{\rm MeV}- 1 \,{\rm GeV})$,
corresponds to a mixing $U_{\nu N}\sim (10^{-4}-10^{-5})$.

For $m_N$ below the kaon-production threshold, experiments exploiting intense kaon fluxes, such as NA62~\cite{NA62:2020mcv,NA62:2021bji} and T2K~\cite{T2K:2019jwa}, can begin to test the mixing values predicted by Eq.~\eqref{mixI}. In this mass range, 
the sterile neutrinos are sufficiently long-lived that a substantial portion of the seesaw-motivated parameter space is constrained by big bang nucleosynthesis (BBN)~\cite{Boyarsky:2020dzc}, although a viable window remains~\cite{Bondarenko:2021cpc}.
At masses around the GeV scale, 
BBN bounds become much less constraining, making this region particularly apt to be experimentally probed. Its direct exploration remains challenging, however, because the minimal seesaw relation predicts extremely small active--sterile mixings, which even dedicated future facilities such as SHiP are generally expected to only approach, but cannot reach~\cite{SHiP:2018xqw}.

A few variations have been proposed to overcome this difficulty, either by considering neutrino-mass models that allow for larger active--sterile mixing while keeping $m_\nu$ small, as in lepton-number-protected scenarios~\cite{Wyler:1982dd,Mohapatra:1986bd,GonzalezGarcia:1988rw,Akhmedov:1995ip,Akhmedov:1995vm,Malinsky:2009df,Abada:2014vea,Barr:2003nn,Malinsky:2005bi},  or by introducing an additional light sector, such as axion-like particles, whose decays can produce HNLs independently from active--sterile mixing~\cite{Abdullahi:2023gdj}. Here, instead, we show that the seesaw region can be explored while adhering to the minimal seesaw model, once the effect of higher dimensional operators is accounted for.

After decoupling states with mass at the electroweak scale or above, the GeV-scale effective theory for the two sterile neutrinos, $N_1$ and $N_2$, allows (up to active-sterile mixing effects) for one unique dimension-five operator,
a transition dipole interaction, $N_1 \sigma^{\mu\nu} N_2\, F_{\mu\nu}$,
with $F_{\mu\nu}$ the photon field-strength tensor. A small sterile dipole is generated, at one loop, by SM charged states at the electroweak scale \cite{Bertuzzo:2024eds}. An even larger dipole can be generated in ultraviolet completions featuring heavy charged states above the electroweak scale, coupled to the sterile neutrinos, see e.g.~\cite{Aparici:2009oua}.

The presence of an electromagnetic dipole transition between $N_1$ and $N_2$ significantly enriches the phenomenology of sterile neutrinos. In particular, it introduces new production mechanisms through neutral-vector-meson  decays, $V^0\to N_1N_2$ (but not through pseudo-scalar meson decays \cite{Bertuzzo:2024eds}) and it induces the radiative decay $N_2\to N_1\gamma$,
as well as decays mediated by an off-shell photon, $N_2\to N_1\ell^+\ell^-$ and $N_2\to N_1 M^0$, where $M^0$ stands for one or several mesons, whenever these channels are kinematically accessible. These processes do not require active--sterile mixing, while several other additional channels open when both the dipole and the mixing are switched on~\cite{Bertuzzo:2024eds}. In particular, all the above $N_2\to N_1$ transitions are complemented by the analogous $N_{1,2}\to \nu$ transitions.

Unlike the active--sterile mixing, the size of the dipole interaction is not directly related to light-neutrino masses. Consequently, the transition dipole may be sufficiently large to substantially modify the production, decay and detection of sterile neutrinos. 
In particular, the dipole might 
significantly improve the sensitivity to small active--sterile mixing, potentially allowing future facilities to probe the region favoured by the minimal seesaw. 
In this Letter, we pursue this possibility.

\section{Two sterile neutrinos with a transition dipole}
\label{sec:model}

We consider an extension of the Standard Model with two gauge-singlet
Weyl fermions, $N'_1$ and $N'_2$. The most relevant operators to be added to the SM Lagrangian can be written as
\begin{equation}
\begin{aligned}
\mathcal{L}_N ={}&
-\left(
M N_1' N_2'
+\frac{\mu_1}{2}N_1'N_1'
+\frac{\mu_2}{2}N_2'N_2'
+\mathrm{h.c.}
\right)
\\
&+i {N_1'}^\dagger \bar{\sigma}^\mu \partial_\mu N_1'
+i {N_2'}^\dagger \bar{\sigma}^\mu \partial_\mu N_2'
\\
&-\left[
\widetilde H^\dagger L_\alpha
\left(
Y_{\alpha 1}N'_1
+Y_{\alpha 2}N'_2
\right)
+\mathrm{h.c.}
\right]
\\
&+\left(
d\,N_1'\sigma^{\mu\nu}N_2'B_{\mu\nu}
+\mathrm{h.c.}
\right)\, ,
\end{aligned}
\label{LN}
\end{equation}
where we adopted the two-component spinor  conventions of Ref.~\cite{Dreiner:2008tw},
and we used primed sterile-neutrino fields for interaction eigenstates, reserving the unprimed notation for mass eigenstates.
The first line in Eq.~\eqref{LN} corresponds to dimension-three mass terms. The second and third lines include dimension-four kinetic terms and Yukawa couplings.
Here $H$ and $L_\alpha$ are the SM Higgs and lepton doublets, $\widetilde H \equiv i\sigma_2H^*$, and a sum over the lepton-flavour index $\alpha=e,\mu,\tau$ is understood. 

The last line in Eq.~\eqref{LN} contains the dimension-five dipole operator coupling the sterile neutrinos to the hypercharge field-strength tensor $B_{\mu\nu}$, first studied in Refs.~\cite{Aparici:2009fh,Aparici:2009oua}.
The dipole Lorentz structure requires antisymmetry in the two fermion fields, therefore the only non-vanishing coupling is a transitional dipole moment connecting $N_1'$ to $N_2'$; note the dipole coefficient $d$ has mass dimension minus one.
At dimension five, the only other allowed operators that contain sterile neutrinos have the form
$N'_iN'_jH^\dagger H$ for $i,j=1,2$: they simply shift the sterile-neutrino masses after electroweak symmetry breaking, and they induce interactions with the Higgs boson. In the following we will consider only the electromagnetic dipole, neglecting operators involving the $Z$ and Higgs bosons, as well as operators with dimension larger than five, since they are not relevant for the phenomenology considered in this work.

The size of the transition dipole can be estimated from the scale
and coupling strength of the ultraviolet completion~\cite{Aparici:2009oua}. Since dipole
operators are generically generated at loop level, naive
dimensional analysis gives
\begin{equation}
    d \sim \frac{g'}{16\pi^2}\frac{g_*^2}{m_*}\,,
    \label{eq:dipole-power-counting}
\end{equation}
where $g'$ is the hypercharge gauge coupling, $g_*$ denotes a
typical coupling between the sterile sector and the UV states
generating the operator, and $m_*$ is their characteristic mass
scale. We use this estimate to define two representative
benchmarks, $g_*=1$ for a weakly coupled UV completion and
$g_*=4\pi$ for a strongly coupled one,
with $m_*=1~\mathrm{TeV}$ as reference value.

After electroweak symmetry breaking, the neutrino mass matrix, in the basis
$(\nu'_e,\nu'_\mu,\nu'_\tau,N'_1,N'_2)$, can be written  as
\begin{equation}
\label{eq:mass_matrix}
\mathcal{M}
=
\begin{pmatrix}
0_\nu & m_D \\
m_D^T & M_N
\end{pmatrix},
\end{equation}
where $0_\nu$ stands for a vanishing $3\times 3$ block, $m_D$ is a $3\times2$ Dirac mass matrix with entries
$(m_D)_{\alpha i}=\frac{v}{\sqrt{2}}Y_{\alpha i}$, where $v\simeq 246$ GeV~\cite{ParticleDataGroup:2026mpi} is the Higgs vacuum expectation value, while $M_N$ is a $2\times 2$ Majorana mass matrix with the structure
\begin{equation}
M_N=
\begin{pmatrix}
\mu_1 & M \\
M & \mu_2
\end{pmatrix} \,.
\label{eq:sterile_mass_matrix}
\end{equation}
The full neutrino mass matrix $\mathcal{M}$ can be diagonalised by a unitary matrix $U$, defined by 
$U^T \mathcal{M} U \equiv \mathcal{M}_{diag}\equiv \text{diag}(m_{\nu_1},m_{\nu_2},m_{\nu_3},m_{N_1},m_{N_2})$. We order the HNL masses according to $m_{N_1} \le m_{N_2}$ and define the relative mass splitting as $\delta \equiv (m_{N_2}-m_{N_1})/(m_{N_2}+m_{N_1})$.

In the seesaw limit, the light-neutrino mass matrix (in the flavour basis) is given by
\begin{eqnarray}
    (m_\nu)_{\alpha\beta} &\simeq& - (m_D M_N^{-1}m_D^T)_{\alpha\beta} \nonumber \\ 
    & \simeq & - \sum_{i=1}^{2}
m_{N_i}\,
U_{\nu_\alpha N_i}U_{\nu_\beta N_i} \,.
\end{eqnarray}
We focus on normal hierarchy of light neutrino masses, with $m_{\nu_1}=0$.~\footnote{Normal ordering is currently
preferred by global fits to neutrino oscillation
data~\cite{Esteban:2024eli}, although inverted
ordering remains experimentally allowed and could be treated similarly.} 
As motivated in Appendix~\ref{App:benchmark_and_seesaw_band},
for the mixing of active neutrinos with $N_1$ we
adopt the simple $0:1:1$ flavour
benchmark, that is, we fix
$U_{\nu_e N_1}=0$ and
$U_{\nu_\mu N_1}=U_{\nu_\tau N_1}$.
The corresponding, approximate seesaw band is given by
\begin{equation}
 \frac{m_{\nu_2}}{2m_{N_1}}\lesssim
 |U_{\nu_{\mu,\tau} N_1}|^2
 \lesssim\frac{m_{\nu_3}}{2m_{N_1}}\,.
 \label{eq:seesaw_band_model}
\end{equation}
The mixing parameters $U_{\nu_\alpha N_2}$ are not fixed by this benchmark
and must be chosen consistently with neutrino oscillation data.
Their flavour structure can differ from that of $N_1$, although their overall sizes are expected to be similar, see the discussion after Eq.~\eqref{eq:CI_mixing}. For simplicity, we assume $U_{\nu_\alpha N_2}=U_{\nu_\alpha N_1}$ throughout our work. As discussed in Sec.~\ref{sec:monophoton}, the specific values of  $U_{\nu_\alpha N_2}$ have little impact on our results since, in the relevant region of parameters, $N_2$ decays promptly through the dipole interaction, before reaching the detector.

\section{Catching sterile neutrinos by the dipole}
\label{sec:monophoton}

The dipole modifies both the production and detection of sterile neutrinos. While conventional production relies on active--sterile mixing through meson decays, such as those of $\pi$, $K$, $D$, $B$ mesons, the dipole allows sterile-neutrino pairs to be produced directly through neutral-vector-meson decays, $V^0\to N_1N_2$, independently of their mixing with active neutrinos. This new channel, with $V^0=\rho^0$, $\omega$, $\phi$, $J/\psi,\Upsilon$, can substantially enhance the sterile-neutrino flux at beam-dump experiments. Details on the production rates and meson yields are provided in Appendix~\ref{app:widths_production}.

Let us now discuss the perspectives for detecting the photons coming from the electromagnetic-dipole interaction of sterile neutrinos. Besides the standard sterile-neutrino decays mediated by
active--sterile mixing (including leptonic channels such as $N_i \to \ell_\alpha^\pm \ell_\beta^\mp \nu$ and semileptonic channels such as $N_i \to \ell_\alpha^\pm \pi^\mp$~\cite{Atre:2009rg,Bondarenko:2018ptm}),
the presence of a dipole opens additional decay channels.
As studied in Refs.~\cite{Barducci:2022gdv,Barducci:2024nvd,Bertuzzo:2024eds},
the dipole induces the radiative decay
$N_2\to N_1\gamma$.
For sufficiently large dipole coefficient, this decay dominates
the width of $N_2$, making it short-lived, so that $N_2$ decays
before reaching the detector, leaving $N_1$
as the relevant long-lived particle. 

In the presence of active--sterile mixing, the
transition dipole also induces effective dipole
interactions between sterile and active neutrinos.
Consequently, $N_1$ can decay through
\begin{equation}
N_1\to\nu\gamma,
\qquad
N_1\to\nu\gamma^*\to\nu\ell^+\ell^-,
\label{eq:dipole_N1_decays}
\end{equation}
where $\ell=e,\mu$ and $\gamma^*$ denotes an off-shell
photon. Unlike $N_2\to N_1\gamma$, these decays
require non-zero active--sterile mixing, which makes $N_1$ much longer lived than $N_2$.
The off-shell decay is suppressed with respect to
the radiative channel by an additional electromagnetic
coupling and three-body phase space, but is enhanced by a collinear divergence in the limit of vanishing $m_\ell$, see Eq.~\eqref{eq:Gamma_N1tonue+e-}. For $m_{N_1}\sim1~\mathrm{GeV}$, this corresponds
to a suppression of order $10^{-2}$ for electrons,
and $10^{-3}$ for muons.

Despite such suppression, the experimental
detection of off-shell photons turns out to be considerably easier. In particular, SHiP is designed to detect the
charged-particle final states with expected negligible  backgrounds~\cite{SHiP:2021nfo}
and T2K already counted charged-lepton pairs with no background~\cite{T2K:2019hnl}. 

A monophoton signal, on the other hand, is particularly challenging
in experiments such as SHiP, where the evacuated
decay volume provides essentially no material for
photon conversion into charged tracks. Although
photons may be detected in the downstream calorimeter,
a single photon does not provide a reconstructible
displaced decay vertex, making background discrimination
particularly challenging. Similarly, the gaseous Time Projection Chambers (TPCs)
of T2K contain little material: the radiation length of
argon gas at atmospheric pressure is of order $10^2$ m,
corresponding to a photon-conversion probability of
order $1\%$ over a metre of gas~\cite{ParticleDataGroup:2026mpi,T2K:2019jwa}.

Alternative strategies for probing the dipole through photon
conversion in argon TPCs are discussed in
Appendix~\ref{app:argon_tpcs}.
In the remainder of this work, we consider only the off-shell
contribution to the dipole-mediated signal,
neglecting $N_1\to\nu\gamma$ in the signal event rate.
Of course, all decay channels are included when computing
the total width of $N_1$.

\begin{figure*}[t]
    \centering
    \includegraphics[width=0.48\textwidth]
    {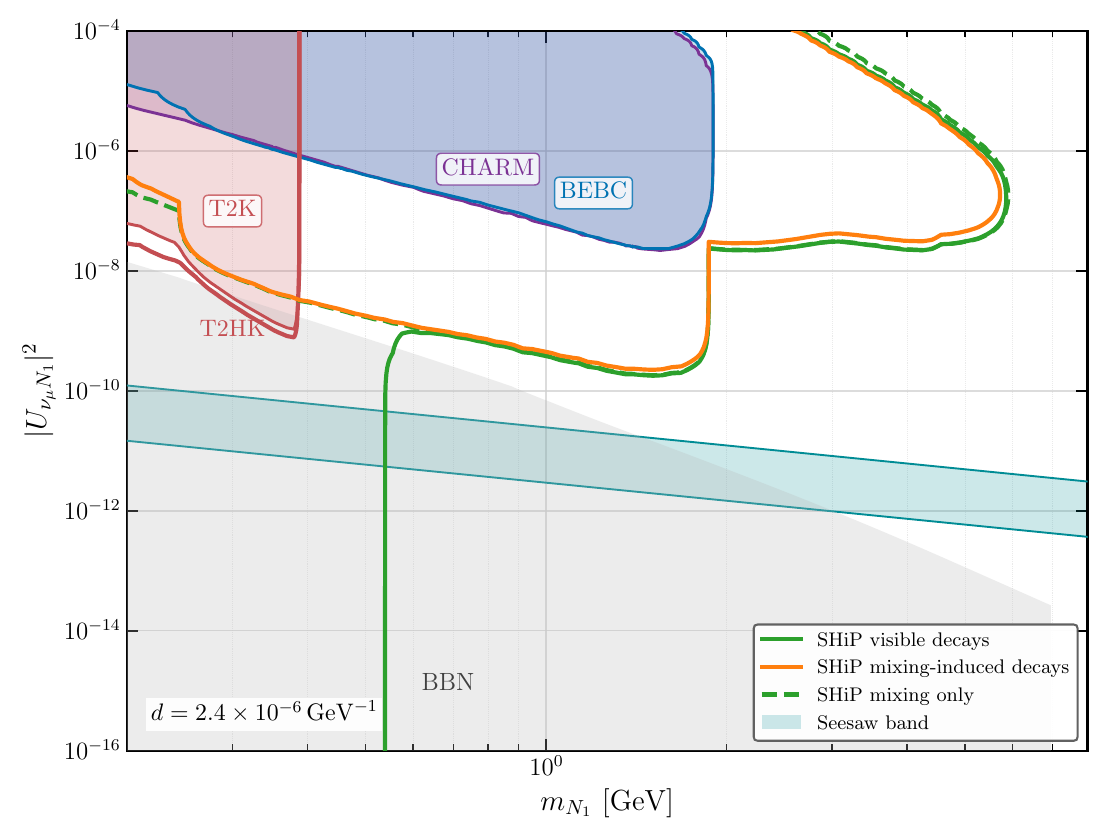}
    \hfill
    \includegraphics[width=0.48\textwidth]{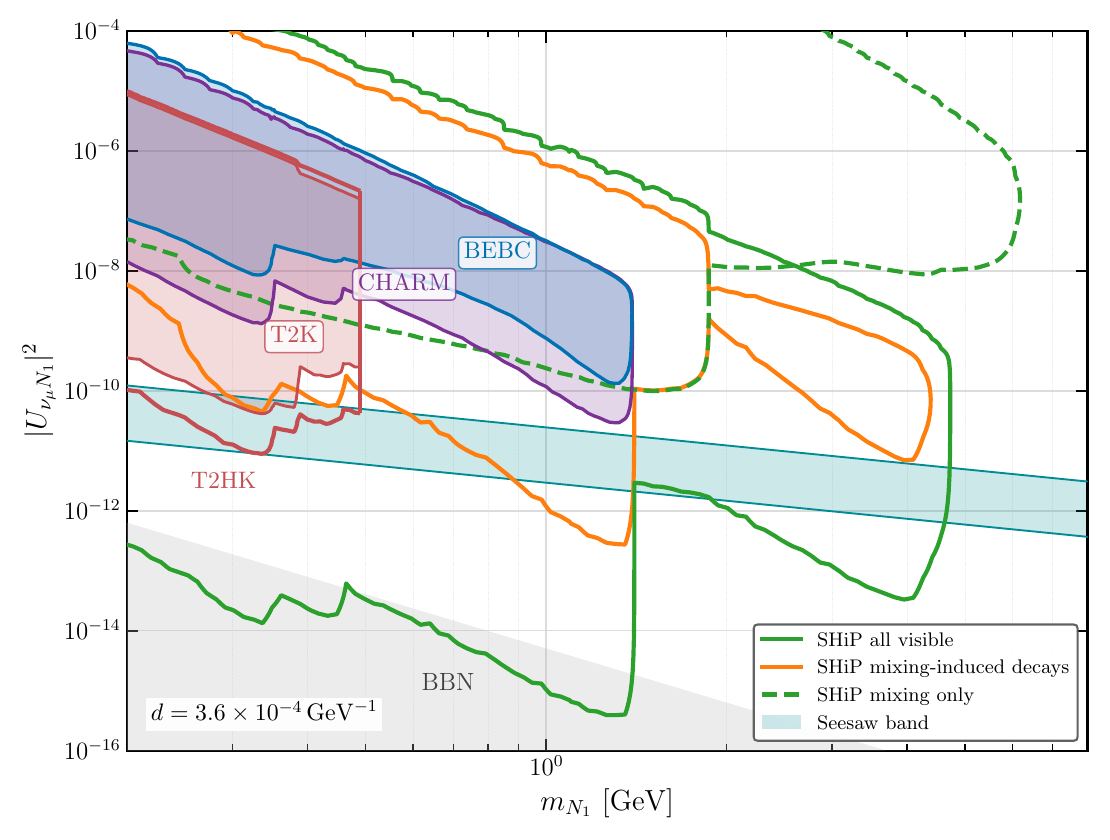}
    \caption{
    Current constraints and projected SHiP sensitivities
    in the $(m_{N_1},|U_{\nu_{\mu}N_1}|^2)$ 
    plane, for a sterile dipole $d = 2.4 \times 10^{-6}~\mathrm{GeV}^{-1}$ (left) and $d=3.6\times10^{-4}\,\mathrm{GeV}^{-1}$ (right),
    with relative mass splitting $\delta=0.1$ and assuming
    $U_{\nu_e N_1}=0$ and $U_{\nu_\mu N_1}=U_{\nu_\tau N_1}$.
    The solid green curve includes all decay channels,
    while the solid orange curve includes only mixing-induced channels (see text). The green dashed curve gives the sensitivity in the zero dipole limit.
    Current constraints from T2K, CHARM, and BEBC are shown
    as shaded regions, and the red line shows the projected T2HK reach. The light-blue band denotes the
    minimal seesaw prediction, while the
    grey region corresponds to
    $\tau_{N_1}>0.02\,\mathrm{s}$, which is disfavoured by BBN~\cite{Boyarsky:2020dzc}.}
    \label{fig:mixing-large-dipole}
\end{figure*}

\section{Probing the minimal seesaw 
}
\label{sec:seesaw}

We now investigate whether a sizeable 
sterile-neutrino dipole $d$ can give experimental access to
the small active--sterile mixing $U_{\nu N}$ predicted by the minimal seesaw. We consider the current constraints
from T2K, CHARM and BEBC, together with the projected
sensitivity of T2HK and SHiP. For T2K/T2HK, we follow Refs.~\cite{T2K:2019jwa,Hyper-Kamiokande:2025fci}.
Further details on the experimental
geometries, flux simulations, and event-selection criteria
are provided in Appendix~\ref{app:widths_production}.

In Fig.~\ref{fig:mixing-large-dipole},
we show the current constraints and the projected SHiP sensitivity
in the plane $(m_{N_1},|U_{\nu_{\mu}N_1}|^2)$, for
the benchmark values $\delta=0.1$~\footnote{The sensitivity would be slightly better for degenerate sterile neutrinos, $\delta=0$, but this requires an additional tuning or symmetry. In the opposite limit with large mass splitting, $\delta\to 1$, $N_2$ decouples and the dipole effects are suppressed.}
and $d = 2.4\times 10^{-6}~\mathrm{GeV}^{-1}$ (left panel) and $d=3.6\times10^{-4}~\mathrm{GeV}^{-1}$ (right panel), corresponding to  weakly and strongly-coupled TeV-scale new physics, respectively.

The solid green curve takes into account all visible decay channels, while the
solid orange curve includes those observable final states which can only be generated by mixing. For the latter, we  conservatively restrict to channels with simple topologies (specifically, $N_1\to\nu e^\pm\mu^\mp$ and $N_1\to\mu^\pm\pi^\mp$), for which high detection efficiencies can reasonably be expected.  
For reference, the green dashed line shows the
standard mixing-only sensitivity, i.e.~the case $d=0$, where we conservatively included only the same subset of decay channels as for the orange line.
To derive the SHiP projected sensitivities, we assume a
background-free search and require 2.44 signal events,
corresponding to the $90\%$ C.L. Feldman--Cousins upper
limit~\cite{Feldman:1997qc}. A background-free search is
well motivated for SHiP, where decays into pairs of charged
particles are expected to provide a very clean signature
and the total background is expected to be reduced to
negligible levels~\cite{SHiP:2021nfo}.

In the left panel, the dipole is too small to drastically change the sensitivity region for masses $m_{N_1} \gtrsim 500~\mathrm{MeV}$. For smaller masses, on the contrary, the sensitivity extends to arbitrarily small mixing, because in this region the signal is dominated by the $N_2 \to N_1 \ell^+ \ell^-$ decay, which is induced by the dipole only. 

In contrast, in the right panel, a large dipole 
extends the
sensitivity towards much smaller active--sterile mixing for any value of $m_{N_1}$.
When all
visible channels are included, SHiP can probe mixings as small
as $|U_{\nu_{\mu}N_1}|^2\sim10^{-15}$  around
$m_{N_1}\sim1~\mathrm{GeV}$, well below the minimal seesaw band. 

In our model, the signal is generated by the combined effect of dipole and active-sterile mixing.
However, alternative scenarios may produce a similar signal. 
Notice in particular that decay channels such as $N_1\to\nu\ell^+\ell^-$, which have the largest branching ratio, contain missing energy carried by the neutrino, and therefore the mass of $N_1$ cannot be fully reconstructed.
Now, consider a scenario with zero mixing,
where the sterile neutrinos can be seen as dark fermions coupled to the SM via the dipole portal only.\footnote{
This corresponds to a model of inelastic dark matter
\cite{Tsai:2019buq,Barducci:2022gdv,Barducci:2024nvd}. 
} 
In this case the charged-lepton-pair signal can be generated by the dipole-only
transition $N_2\to N_1\ell^+\ell^-$, since the same number of events could be obtained for some appropriate values of $m_{N_1}$, $\delta$ and $d$.

\begin{figure*}[t]
    \centering
    \includegraphics[width=0.48\textwidth]
    {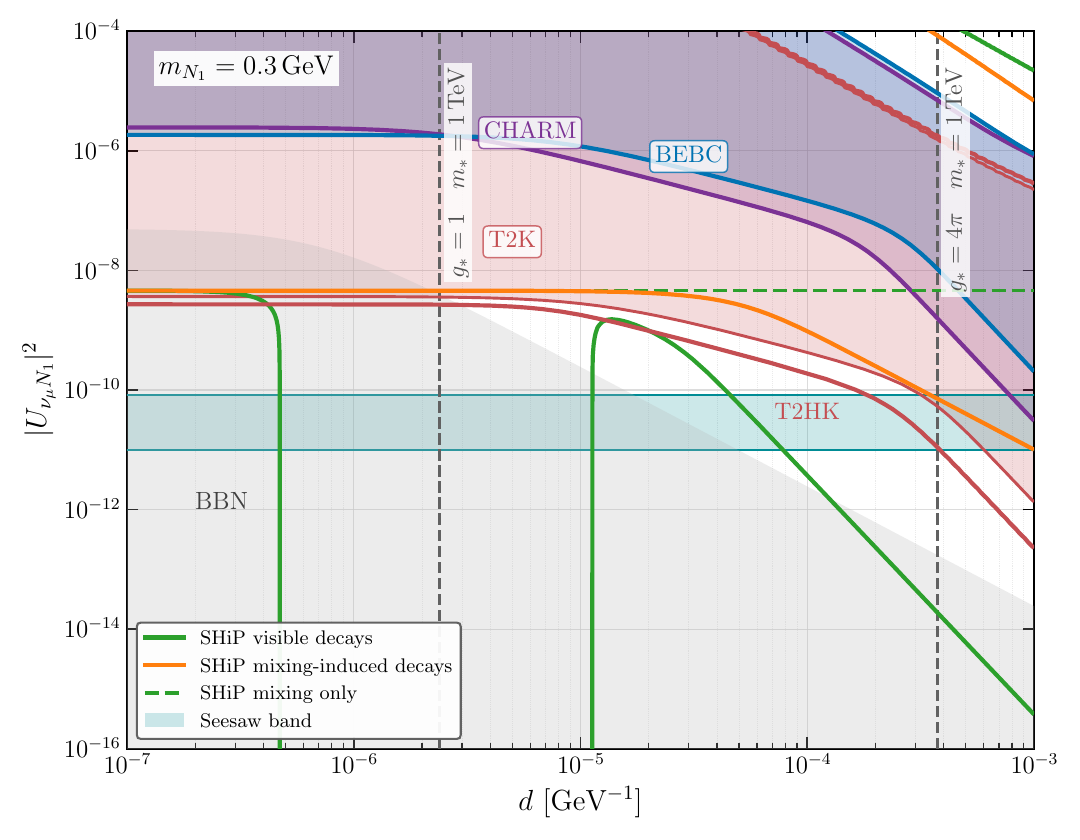}
    \hfill
    \includegraphics[width=0.48\textwidth]
    {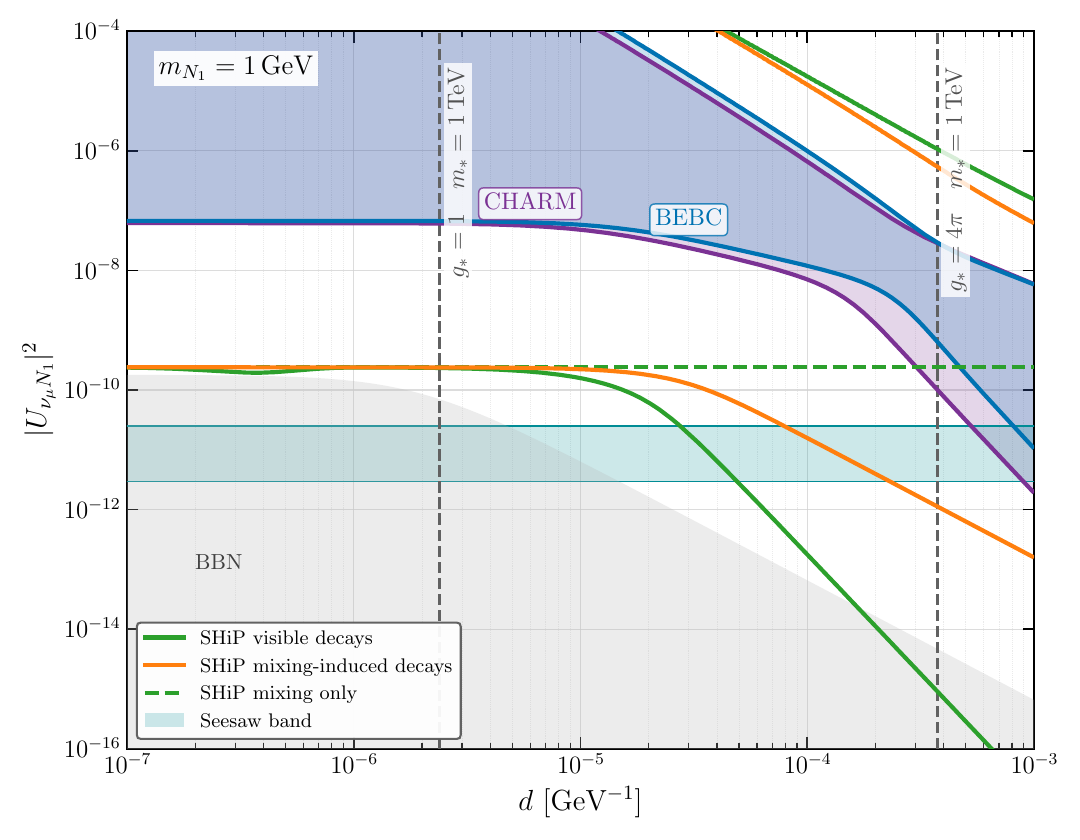}
    \caption{
    Current constraints and projected SHiP sensitivities in the
    $(d,|U_{\nu_{\mu}N_1}|^2)$ plane,
    for $N_1$ mass below
    the kaon threshold, $m_{N_1}=0.3\,\mathrm{GeV}$ (left panel),
    and above it, $m_{N_1}=1\,\mathrm{GeV}$~(right panel),
    with sterile mass splitting $\delta=0.1$.
    We adopted the flavour benchmark
    $U_{\nu_e N_1}=0$ and
    $U_{\nu_\mu N_1}=U_{\nu_\tau N_1}$.
    The solid green (orange) curves delimit the sensitivity region including all visible decay channels (only mixing-induced channels, see text). The dashed green 
    curve shows the sensitivity to the mixing in the no-dipole limit,
    for comparison. Current constraints from T2K, CHARM,
    and BEBC are shown as shaded regions, and the projected T2HK sensitivity as a red line.
    The light-blue region denotes the minimal seesaw band,
    while the grey region is
    disfavoured by BBN. The vertical dashed lines mark
    the dipole benchmark values for a new-physics
    scale $m_*=1\,\mathrm{TeV}$ with weak (strong) coupling
    $g_*=1$ ($g_*=4\pi$), corresponding to the left (right) panel of Fig.~\ref{fig:mixing-large-dipole}.
    }
    \label{fig:dipole-mixing-plane}
\end{figure*}

In order to  establish that a non-zero active-sterile mixing is actually at the origin of the signal,
a smoking gun is
provided by
the mixing-induced decay channels, shown by the solid orange
curve in Fig.~\ref{fig:mixing-large-dipole}. These correspond to final states which can only be generated by mixing, such as the flavour-violating channels, $e^\pm\mu^\mp$, and the charged-current semileptonic channels, e.g.~$\mu^{\pm}\pi^{\mp}$.
Even though this subset of decay channels gives a more modest sensitivity, it still
reaches the seesaw band for  $m_{N_1}\sim 1~\mathrm{GeV}$, where SHiP can probe mixing of order
$|U_{\nu_{\mu}N_1}|\sim 10^{-6}$. 
The dipole-plus-mixing scenario can also be distinguished from
the mixing-only case, since in the latter one does not
expect the large hierarchy between the 
dipole-enhanced channels and the mixing-induced ones.

The shaded regions show that the presence of a dipole strengthens existing constraints. It also improves the reach of the future T2K upgrade: T2HK will be able to probe the
seesaw band for $m_{N_1}\lesssim 400$ MeV, with its reach
limited by the kinematic closure of $\phi$ meson production. CHARM and BEBC extend the constraints to larger
masses, but do not touch the seesaw band. A significant
part of the GeV-scale seesaw region remains open and can
be targeted by SHiP.

Let us study how the sensitivities vary as a function of the dipole size.
The combined dependence on $d$ and $|U_{\nu_{\mu}N_1}|^2$
is illustrated explicitly
in Fig.~\ref{fig:dipole-mixing-plane},  for two representative sterile masses, $m_{N_1}=0.3~\mathrm{GeV}$ (left) and $1~\mathrm{GeV}$ (right). 
At small $d$, the sensitivity approaches the conventional mixing-only scenario.\footnote{
Note that, in the presence of active-sterile mixing, $d$ receives an irreducible, calculable contribution from one-loop electroweak diagrams \cite{Bertuzzo:2024eds}. Here this has been neglected, since it is subdominant for $d\gtrsim 5\cdot 10^{-7}$ GeV$^{-1}$. Should a precision study of the region of smaller $d$ become of interest, it would be necessary to include one-loop electroweak corrections to the dipole, as well as to the various decay widths.}
At large $d$, the enhanced sterile production extends the reach towards a progressively smaller mixing. In the left panel, for intermediate values of $d$, the dipole-only channel $N_2\to N_1\ell^+\ell^-$ produces a signal even for vanishing mixing, as already explained above.

For $m_{N_1}$ below the kaon threshold (left panel), T2K is already probing part of the
relevant region, and only a small gap remains between the
T2K bound and the region disfavoured by BBN, that could
be covered by SHiP. If a signal were observed in
SHiP, it would be challenging to attribute it to a seesaw-like mixing, as a dipole-only interpretation would remain possible as well.
For $m_{N_1}\sim 1~\mathrm{GeV}$ (right
panel), the situation is more promising.
We find that, for
$d\gtrsim10^{-5}~\mathrm{GeV}^{-1}$, SHiP should be able
to probe the minimal seesaw band, while for $d\gtrsim 10^{-4}~\mathrm{GeV}^{-1}$ it could be sensitive also to
the decay channels induced by mixing only,
with a chance to distinguish the dipole-plus-mixing
scenario from a pure-dipole interpretation.

\section{Conclusions}
\label{sec:conclusion}

Our analysis led to two main results:
\begin{itemize}
\item[(i)] GeV-scale HNLs can be discovered through their electromagnetic dipole, as they are copiously produced in beam-dump and accelerator-neutrino experiments via meson-photon mixing. Their subsequent decays into
charged-lepton pairs through an off-shell photon provide
a promising experimental signature for
low-density detectors.
\item[(ii)] The HNL dipole can be instrumental to access a tiny mixing between them and active neutrinos, at the level predicted by the minimal seesaw mechanism: we find that T2HK near detector will probe the seesaw band for HNL masses of a few hundred MeV, and SHiP will extend the coverage to a few GeV.
\end{itemize}

The expected SHiP sensitivity to different decay channels could allow to discriminate between dipole-induced processes, and mixing-induced ones.
The discovery of an active-sterile mixing compatible with the minimal seesaw prediction would shed light on the origin of light neutrino masses.

\section{acknowledgments}

We thank Claudio Giganti for his involvement in the early stages of this work and useful discussions.
MF received funding from the European Union’s Horizon Europe research and innovation programme under the Marie Sklodowska-Curie Staff Exchange grant agreement No 101086085 – ASYMMETRY. EB acknowledges the financial support of PRD 2026 funds at UNIMORE, as well as partial support from the Italian INFN program on Theoretical Astroparticle Physics (TAsP). EB \& MF thank ICTP-SAIFR, S\~ao Paulo, for hospitality during the development of this work.

\bibliographystyle{utphys}
\bibliography{Refs}

@article{Minkowski:1977sc,
	author = {Minkowski, Peter},
	doi = {10.1016/0370-2693(77)90435-X},
	journal = {Phys. Lett. B},
	pages = {421--428},
	reportnumber = {Print-77-0182 (BERN)},
	title = {{$\mu \to e\gamma$ at a Rate of One Out of $10^{9}$ Muon Decays?}},
	volume = {67},
	year = {1977}}

@article{Gell-Mann:1979vob,
	archiveprefix = {arXiv},
	author = {Gell-Mann, Murray and Ramond, Pierre and Slansky, Richard},
	eprint = {1306.4669},
	journal = {Conf. Proc. C},
	pages = {315--321},
	primaryclass = {hep-th},
	reportnumber = {PRINT-80-0576},
	title = {{Complex Spinors and Unified Theories}},
	volume = {790927},
	year = {1979}}

@article{Yanagida:1979as,
	author = {Yanagida, Tsutomu},
	editor = {Sawada, Osamu and Sugamoto, Akio},
	journal = {Conf. Proc. C},
	pages = {95--99},
	reportnumber = {KEK-79-18-95},
	title = {{Horizontal gauge symmetry and masses of neutrinos}},
	volume = {7902131},
	year = {1979}}

@article{Mohapatra:1979ia,
	author = {Mohapatra, Rabindra N. and Senjanovic, Goran},
	doi = {10.1103/PhysRevLett.44.912},
	journal = {Phys. Rev. Lett.},
	pages = {912},
	reportnumber = {MDDP-TR-80-060, MDDP-PP-80-105, CCNY-HEP-79-10},
	title = {{Neutrino Mass and Spontaneous Parity Nonconservation}},
	volume = {44},
	year = {1980}}

@article{NA62:2020mcv,
	archiveprefix = {arXiv},
	author = {Cortina Gil, Eduardo and others},
	collaboration = {NA62},
	doi = {10.1016/j.physletb.2020.135599},
	eprint = {2005.09575},
	journal = {Phys. Lett. B},
	pages = {135599},
	primaryclass = {hep-ex},
	reportnumber = {CERN-EP-2020-089},
	title = {{Search for heavy neutral lepton production in $K^+$ decays to positrons}},
	volume = {807},
	year = {2020}}

@article{NA62:2021bji,
	archiveprefix = {arXiv},
	author = {Cortina Gil, Eduardo and others},
	collaboration = {NA62},
	doi = {10.1016/j.physletb.2021.136259},
	eprint = {2101.12304},
	journal = {Phys. Lett. B},
	pages = {136259},
	primaryclass = {hep-ex},
	reportnumber = {CERN-EP-2021-018},
	title = {{Search for $K^+$ decays to a muon and invisible particles}},
	volume = {816},
	year = {2021}}

@article{T2K:2019jwa,
	archiveprefix = {arXiv},
	author = {Abe, K. and others},
	collaboration = {T2K},
	doi = {10.1103/PhysRevD.100.052006},
	eprint = {1902.07598},
	journal = {Phys. Rev. D},
	number = {5},
	pages = {052006},
	primaryclass = {hep-ex},
	title = {{Search for heavy neutrinos with the T2K near detector ND280}},
	volume = {100},
	year = {2019}}

@article{Boyarsky:2020dzc,
	archiveprefix = {arXiv},
	author = {Boyarsky, Alexey and Ovchynnikov, Maksym and Ruchayskiy, Oleg and Syvolap, Vsevolod},
	doi = {10.1103/PhysRevD.104.023517},
	eprint = {2008.00749},
	journal = {Phys. Rev. D},
	number = {2},
	pages = {023517},
	primaryclass = {hep-ph},
	title = {{Improved big bang nucleosynthesis constraints on heavy neutral leptons}},
	volume = {104},
	year = {2021}}

@article{Bondarenko:2021cpc,
	archiveprefix = {arXiv},
	author = {Bondarenko, Kyrylo and Boyarsky, Alexey and Klaric, Juraj and Mikulenko, Oleksii and Ruchayskiy, Oleg and Syvolap, Vsevolod and Timiryasov, Inar},
	doi = {10.1007/JHEP07(2021)193},
	eprint = {2101.09255},
	journal = {JHEP},
	pages = {193},
	primaryclass = {hep-ph},
	title = {{An allowed window for heavy neutral leptons below the kaon mass}},
	volume = {07},
	year = {2021}}

@article{SHiP:2018xqw,
	archiveprefix = {arXiv},
	author = {Ahdida, C. and others},
	collaboration = {SHiP},
	doi = {10.1007/JHEP04(2019)077},
	eprint = {1811.00930},
	journal = {JHEP},
	pages = {077},
	primaryclass = {hep-ph},
	title = {{Sensitivity of the SHiP experiment to Heavy Neutral Leptons}},
	volume = {04},
	year = {2019}}

@article{Mohapatra:1986bd,
	author = {Mohapatra, R.N. and Valle, J.W.F.},
	doi = {10.1103/PhysRevD.34.1642},
	journal = {Phys. Rev. D},
	pages = {1642--1645},
	title = {{Neutrino Mass and Baryon Number Nonconservation in Superstring Models}},
	volume = {34},
	year = {1986}}

@article{Malinsky:2005bi,
	archiveprefix = {arXiv},
	author = {Malinsky, Michal and Romao, J.C. and Valle, J.W.F.},
	doi = {10.1103/PhysRevLett.95.161801},
	eprint = {hep-ph/0506296},
	journal = {Phys. Rev. Lett.},
	pages = {161801},
	title = {{Novel Supersymmetric SO(10) Seesaw Mechanism}},
	volume = {95},
	year = {2005}}

@article{Aparici:2009fh,
	archiveprefix = {arXiv},
	author = {Aparici, Alberto and Kim, Kyungwook and Santamaria, Arcadi and Wudka, Jose},
	doi = {10.1103/PhysRevD.80.013010},
	eprint = {0904.3244},
	journal = {Phys. Rev. D},
	pages = {013010},
	primaryclass = {hep-ph},
	title = {{Right-handed neutrino magnetic moments}},
	volume = {80},
	year = {2009}}

@article{Aparici:2009oua,
	archiveprefix = {arXiv},
	author = {Aparici, Alberto and Santamaria, Arcadi and Wudka, Jose},
	doi = {10.1088/0954-3899/37/7/075012},
	eprint = {0911.4103},
	journal = {J. Phys. G},
	pages = {075012},
	primaryclass = {hep-ph},
	title = {{A model for right-handed neutrino magnetic moments}},
	volume = {37},
	year = {2010}}

@article{Bertuzzo:2024eds,
	archiveprefix = {arXiv},
	author = {Bertuzzo, Enrico and Frigerio, Michele},
	doi = {10.21468/SciPostPhys.18.4.140},
	eprint = {2412.10101},
	journal = {SciPost Phys.},
	number = {4},
	pages = {140},
	primaryclass = {hep-ph},
	title = {{Two portals to GeV sterile neutrinos: Dipole versus mixing}},
	volume = {18},
	year = {2025}}

@article{Coloma:2023oxx,
	archiveprefix = {arXiv},
	author = {Coloma, Pilar and Mart{\'\i}n-Albo, Justo and Urrea, Salvador},
	doi = {10.1103/PhysRevD.109.035013},
	eprint = {2309.06492},
	journal = {Phys. Rev. D},
	number = {3},
	pages = {035013},
	primaryclass = {hep-ph},
	reportnumber = {IFT-UAM/CSIC-23-111, IFIC/23-40, FTUV-23-0823.4331},
	title = {{Discovering long-lived particles at DUNE}},
	volume = {109},
	year = {2024}}

@article{DUNE:2021tad,
	archiveprefix = {arXiv},
	author = {Abed Abud, A. and others},
	collaboration = {DUNE},
	doi = {10.3390/instruments5040031},
	eprint = {2103.13910},
	journal = {Instruments},
	number = {4},
	pages = {31},
	primaryclass = {physics.ins-det},
	title = {{Deep Underground Neutrino Experiment (DUNE) Near Detector Conceptual Design Report}},
	volume = {5},
	year = {2021}}

@article{CooperSarkar:1985nh,
	author = {Cooper-Sarkar, A. M. and others},
	collaboration = {WA66},
	doi = {10.1016/0370-2693(85)91493-5},
	journal = {Phys. Lett. B},
	pages = {207--211},
	title = {{Search for Heavy Neutrino Decays in the BEBC Beam Dump Experiment}},
	volume = {160},
	year = {1985}}

@article{Dreiner:2008tw,
	archiveprefix = {arXiv},
	author = {Dreiner, Herbi K. and Haber, Howard E. and Martin, Stephen P.},
	doi = {10.1016/j.physrep.2010.05.002},
	eprint = {0812.1594},
	journal = {Phys. Rept.},
	pages = {1--196},
	primaryclass = {hep-ph},
	reportnumber = {BN-TH-2008-12, SCIPP-08-08, FERMILAB-PUB-09-855-T, BN-TH-2008-12 and SCIPP-08/08},
	title = {{Two-component spinor techniques and Feynman rules for quantum field theory and supersymmetry}},
	volume = {494},
	year = {2010}}

@article{Esteban:2024eli,
	archiveprefix = {arXiv},
	author = {Esteban, Ivan and Gonzalez-Garcia, M. C. and Maltoni, Michele and Martinez-Soler, Ivan and Pinheiro, Jo{\~a}o Paulo and Schwetz, Thomas},
	doi = {10.1007/JHEP12(2024)216},
	eprint = {2410.05380},
	journal = {JHEP},
	pages = {216},
	primaryclass = {hep-ph},
	reportnumber = {IFT-UAM/CSIC-24-140, YITP-SB-2024-24, IPPP/24/64, IPPP/24/64, IFT-UAM/CSIC-24-140, YITP-SB-2024-24},
	title = {{NuFit-6.0: updated global analysis of three-flavor neutrino oscillations}},
	volume = {12},
	year = {2024}}

@article{Abada:2014vea,
	archiveprefix = {arXiv},
	author = {Abada, Asmaa and Lucente, Michele},
	doi = {10.1016/j.nuclphysb.2014.06.003},
	eprint = {1401.1507},
	journal = {Nucl. Phys. B},
	pages = {651--678},
	primaryclass = {hep-ph},
	reportnumber = {LPT-ORSAY-13-39, SISSA-59-2013-FISI},
	title = {{Looking for the minimal inverse seesaw realisation}},
	volume = {885},
	year = {2014}}

@article{Casas:2001sr,
	archiveprefix = {arXiv},
	author = {Casas, J. A. and Ibarra, A.},
	doi = {10.1016/S0550-3213(01)00475-8},
	eprint = {hep-ph/0103065},
	journal = {Nucl. Phys. B},
	pages = {171--204},
	title = {{Oscillating neutrinos and muon to e, gamma}},
	volume = {618},
	year = {2001}}

@article{Moffat:2017feq,
	archiveprefix = {arXiv},
	author = {Moffat, K. and Pascoli, S. and Weiland, C.},
	eprint = {1712.07611},
	primaryclass = {hep-ph},
	title = {{Equivalence between massless neutrinos and lepton number conservation in fermionic singlet extensions of the Standard Model}},
	year = {2017}}

@article{SHiP:2021nfo,
	archiveprefix = {arXiv},
	author = {Ahdida, C. and others},
	collaboration = {SHiP},
	doi = {10.1140/epjc/s10052-022-10346-5},
	eprint = {2112.01487},
	journal = {Eur. Phys. J. C},
	number = {5},
	pages = {486},
	primaryclass = {physics.ins-det},
	title = {{The SHiP experiment at the proposed CERN SPS Beam Dump Facility}},
	volume = {82},
	year = {2022}}

@article{Feldman:1997qc,
	archiveprefix = {arXiv},
	author = {Feldman, Gary J. and Cousins, Robert D.},
	doi = {10.1103/PhysRevD.57.3873},
	eprint = {physics/9711021},
	journal = {Phys. Rev. D},
	pages = {3873--3889},
	reportnumber = {HUTP-97-A096},
	title = {{A Unified approach to the classical statistical analysis of small signals}},
	volume = {57},
	year = {1998}}

@article{Drewes:2022akb,
	archiveprefix = {arXiv},
	author = {Drewes, Marco and Klari{\'c}, Juraj and L{\'o}pez-Pav{\'o}n, Jacobo},
	doi = {10.1140/epjc/s10052-022-11100-7},
	eprint = {2207.02742},
	journal = {Eur. Phys. J. C},
	number = {12},
	pages = {1176},
	primaryclass = {hep-ph},
	title = {{New benchmark models for heavy neutral lepton searches}},
	volume = {82},
	year = {2022}}

@article{ParticleDataGroup:2026mpi,
	author = {Takahashi, F. and others},
	collaboration = {Particle Data Group},
	doi = {10.1142/s0217751x26300115},
	journal = {Int. J. Mod. Phys. A},
	number = {22},
	pages = {2630011},
	title = {{Review of Particle Physics}},
	volume = {41},
	year = {2026}}

@article{Atre:2009rg,
	archiveprefix = {arXiv},
	author = {Atre, Anupama and Han, Tao and Pascoli, Silvia and Zhang, Bin},
	doi = {10.1088/1126-6708/2009/05/030},
	eprint = {0901.3589},
	journal = {JHEP},
	pages = {030},
	primaryclass = {hep-ph},
	title = {{The Search for Heavy Majorana Neutrinos}},
	volume = {05},
	year = {2009}}

@article{Bondarenko:2018ptm,
	archiveprefix = {arXiv},
	author = {Bondarenko, Kyrylo and Boyarsky, Alexey and Gorbunov, Dmitry and Ruchayskiy, Oleg},
	doi = {10.1007/JHEP11(2018)032},
	eprint = {1805.08567},
	journal = {JHEP},
	pages = {032},
	primaryclass = {hep-ph},
	title = {{Phenomenology of GeV-scale Heavy Neutral Leptons}},
	volume = {11},
	year = {2018}}

@article{Barducci:2022gdv,
	archiveprefix = {arXiv},
	author = {Barducci, Daniele and Bertuzzo, Enrico and Taoso, Marco and Toni, Claudio},
	doi = {10.1007/JHEP03(2023)239},
	eprint = {2209.13469},
	journal = {JHEP},
	pages = {239},
	primaryclass = {hep-ph},
	title = {{Probing right-handed neutrinos dipole operators}},
	volume = {03},
	year = {2023}}

@article{Barducci:2024nvd,
	archiveprefix = {arXiv},
	author = {Barducci, D. and Bertuzzo, E. and Taoso, M. and Ternes, C. A. and Toni, C.},
	doi = {10.1007/JHEP10(2024)016},
	eprint = {2406.17599},
	journal = {JHEP},
	pages = {016},
	primaryclass = {hep-ph},
	title = {{Illuminating the dark: mono-{\ensuremath{\gamma}} signals at NA62}},
	volume = {10},
	year = {2024}}

@article{Coloma:2020lgy,
	archiveprefix = {arXiv},
	author = {Coloma, Pilar and Fern{\'a}ndez-Mart{\'\i}nez, Enrique and Gonz{\'a}lez-L{\'o}pez, Manuel and Hern{\'a}ndez-Garc{\'\i}a, Josu and Pavlovic, Zarko},
	doi = {10.1140/epjc/s10052-021-08861-y},
	eprint = {2007.03701},
	journal = {Eur. Phys. J. C},
	number = {1},
	pages = {78},
	primaryclass = {hep-ph},
	reportnumber = {FERMILAB-PUB-20-269-ND},
	title = {{GeV-scale neutrinos: interactions with mesons and DUNE sensitivity}},
	volume = {81},
	year = {2021}}

@article{Beacham:2019nyx,
	archiveprefix = {arXiv},
	author = {Beacham, J. and others},
	doi = {10.1088/1361-6471/ab4cd2},
	eprint = {1901.09966},
	journal = {J. Phys. G},
	number = {1},
	pages = {010501},
	primaryclass = {hep-ex},
	reportnumber = {CERN-PBC-REPORT-2018-007},
	title = {{Physics Beyond Colliders at CERN: Beyond the Standard Model Working Group Report}},
	volume = {47},
	year = {2020}}

@article{Abdullahi:2023gdj,
	archiveprefix = {arXiv},
	author = {Abdullahi, Asli M. and de Gouv{\^e}a, Andr{\'e} and Dutta, Bhaskar and Shoemaker, Ian M. and Tabrizi, Zahra},
	doi = {10.1103/PhysRevLett.133.261802},
	eprint = {2311.07713},
	journal = {Phys. Rev. Lett.},
	number = {26},
	pages = {261802},
	primaryclass = {hep-ph},
	reportnumber = {IFT-UAM/CSIC-23-145, MI-HET-820, FERMILAB-PUB-23-747-T},
	title = {{Heavy Neutral Leptons via Axionlike Particles at Neutrino Facilities}},
	volume = {133},
	year = {2024}}

@article{Wyler:1982dd,
	author = {Wyler, D. and Wolfenstein, L.},
	doi = {10.1016/0550-3213(83)90482-0},
	journal = {Nucl. Phys. B},
	pages = {205--214},
	reportnumber = {CERN-TH-3435},
	title = {{Massless Neutrinos in Left-Right Symmetric Models}},
	volume = {218},
	year = {1983}}

@article{GonzalezGarcia:1988rw,
	author = {Gonz\'alez-Garc\'ia, M. C. and Valle, J. W. F.},
	doi = {10.1016/0370-2693(89)91131-3},
	journal = {Phys. Lett. B},
	pages = {360--366},
	reportnumber = {CERN-TH-5170-88, FTUV-10-88},
	title = {{Fast Decaying Neutrinos and Observable Flavor Violation in a New Class of Majoron Models}},
	volume = {216},
	year = {1989}}

@article{Akhmedov:1995ip,
	archiveprefix = {arXiv},
	author = {Akhmedov, Evgeny K. and Lindner, Manfred and Schnapka, Erhard and Valle, J. W. F.},
	doi = {10.1016/0370-2693(95)01504-3},
	eprint = {hep-ph/9507275},
	journal = {Phys. Lett. B},
	pages = {270--280},
	reportnumber = {IC-95-125, TUM-HEP-221-95, MPI-PHT-95-35, FTUV-95-34, IFIC-95-36},
	title = {{Left-right symmetry breaking in NJL approach}},
	volume = {368},
	year = {1996}}

@article{Akhmedov:1995vm,
	archiveprefix = {arXiv},
	author = {Akhmedov, Evgeny K. and Lindner, Manfred and Schnapka, Erhard and Valle, J. W. F.},
	doi = {10.1103/PhysRevD.53.2752},
	eprint = {hep-ph/9509255},
	journal = {Phys. Rev. D},
	pages = {2752--2780},
	reportnumber = {IC-95-126, TUM-HEP-222-95, MPI-PHT-95-70, FTUV-95-36, IFIC-95-38},
	title = {{Dynamical left-right symmetry breaking}},
	volume = {53},
	year = {1996}}

@article{Malinsky:2009df,
	archiveprefix = {arXiv},
	author = {Malinsky, Michal and Ohlsson, Tommy and Xing, Zhi-zhong and Zhang, He},
	doi = {10.1016/j.physletb.2009.07.038},
	eprint = {0905.2889},
	journal = {Phys. Lett. B},
	pages = {242--248},
	primaryclass = {hep-ph},
	reportnumber = {NORDITA-2009-31},
	title = {{Non-unitary neutrino mixing and CP violation in the minimal inverse seesaw model}},
	volume = {679},
	year = {2009}}

@article{Barr:2003nn,
	archiveprefix = {arXiv},
	author = {Barr, S. M.},
	doi = {10.1103/PhysRevLett.92.101601},
	eprint = {hep-ph/0309152},
	journal = {Phys. Rev. Lett.},
	pages = {101601},
	title = {{A Different seesaw formula for neutrino masses}},
	volume = {92},
	year = {2004}}

@article{Bergsma:1986is,
	author = {Dorenbosch, J. and others},
	collaboration = {CHARM},
	doi = {10.1016/0370-2693(86)91601-1},
	journal = {Phys. Lett. B},
	pages = {473--478},
	title = {{A Search for Decays of Heavy Neutrinos in the Mass Range 0.5-GeV to 2.8-GeV}},
	volume = {166},
	year = {1986}}

@article{Boiarska:2021charm,
	archiveprefix = {arXiv},
	author = {Boiarska, Iryna and Boyarsky, Alexey and Mikulenko, Oleksii and Ovchynnikov, Maksym},
	doi = {10.1103/PhysRevD.104.095019},
	eprint = {2107.14685},
	journal = {Phys. Rev. D},
	pages = {095019},
	primaryclass = {hep-ph},
	title = {{Constraints from the CHARM experiment on heavy neutral leptons with tau mixing}},
	volume = {104},
	year = {2021}}

@article{DUNE:2024wvj,
	archiveprefix = {arXiv},
	author = {Abed Abud, Adam and others},
	collaboration = {DUNE},
	doi = {10.1088/1748-0221/19/12/P12005},
	eprint = {2408.12725},
	journal = {JINST},
	number = {12},
	pages = {P12005},
	primaryclass = {physics.ins-det},
	reportnumber = {FERMILAB-TM-2833-LBNF},
	title = {{DUNE Phase~II: scientific opportunities, detector concepts, technological solutions}},
	volume = {19},
	year = {2024}}

@article{Tsai:2019buq,
	archiveprefix = {arXiv},
	author = {Tsai, Yu-Dai and deNiverville, Patrick and Liu, Ming Xiong},
	doi = {10.1103/PhysRevLett.126.181801},
	eprint = {1908.07525},
	journal = {Phys. Rev. Lett.},
	number = {18},
	pages = {181801},
	primaryclass = {hep-ph},
	reportnumber = {FERMILAB-PUB-19-393-A-PPD},
	title = {{Dark Photon and Muon $g-2$ Inspired Inelastic Dark Matter Models at the High-Energy Intensity Frontier}},
	volume = {126},
	year = {2021}}

@article{Bierlich:2022pfr,
	archiveprefix = {arXiv},
	author = {Bierlich, Christian and others},
	doi = {10.21468/SciPostPhysCodeb.8},
	eprint = {2203.11601},
	journal = {SciPost Phys. Codebases},
	pages = {8},
	primaryclass = {hep-ph},
	title = {{A comprehensive guide to the physics and usage of PYTHIA 8.3}},
	year = {2022}}

@article{Barouki:2022bebc,
	archiveprefix = {arXiv},
	author = {Barouki, Ryan and Marocco, Giacomo and Sarkar, Subir},
	doi = {10.21468/SciPostPhys.13.5.118},
	eprint = {2208.00416},
	journal = {SciPost Phys.},
	number = {5},
	pages = {118},
	primaryclass = {hep-ph},
	title = {{Blast from the past II: Constraints on heavy neutral leptons from the BEBC WA66 beam dump experiment}},
	volume = {13},
	year = {2022}}

@article{T2K:2019hnl,
	archiveprefix = {arXiv},
	author = {Abe, K. and others},
	collaboration = {T2K},
	doi = {10.1103/PhysRevD.100.052006},
	eprint = {1902.07598},
	journal = {Phys. Rev. D},
	number = {5},
	pages = {052006},
	primaryclass = {hep-ex},
	title = {{Search for heavy neutrinos with the T2K near detector ND280}},
	volume = {100},
	year = {2019}}

@article{Ferrillo:2023hhg,
	archiveprefix = {arXiv},
	author = {Ferrillo, Martina and Ovchynnikov, Maksym and Resnati, Filippo and De Roeck, Albert},
	doi = {10.1007/JHEP02(2024)196},
	eprint = {2312.14868},
	journal = {JHEP},
	pages = {196},
	primaryclass = {hep-ph},
	title = {{Improving the potential of BDF@SPS to search for new physics with liquid argon time projection chambers}},
	volume = {02},
	year = {2024}}

@article{Hyper-Kamiokande:2025fci,
	archiveprefix = {arXiv},
	author = {Abe, K. and others},
	collaboration = {Hyper-Kamiokande},
	doi = {10.1140/epjc/s10052-025-14938-9},
	eprint = {2505.15019},
	journal = {Eur. Phys. J. C},
	note = {[Erratum: Eur.Phys.J.C 86, 1066 (2026)]},
	number = {2},
	pages = {170},
	primaryclass = {hep-ex},
	title = {{Sensitivity of the Hyper-Kamiokande experiment to neutrino oscillation parameters using accelerator neutrinos}},
	volume = {86},
	year = {2026}}

@article{Pauli:1941zz,
	author = {Pauli, W.},
	doi = {10.1103/RevModPhys.13.203},
	journal = {Rev. Mod. Phys.},
	pages = {203--232},
	title = {{Relativistic Field Theories of Elementary Particles}},
	volume = {13},
	year = {1941}}


\clearpage
\appendix

\section{Active-sterile mixing and seesaw band}
\label{App:benchmark_and_seesaw_band}

In this appendix, we motivate the flavour structure adopted
for the active--sterile mixing and derive the corresponding
seesaw band, used in our phenomenological study.

Under the seesaw condition  $|\epsilon_{\alpha i}|\ll 1$, where
$\epsilon\equiv m_DM_N^{-1}$ is a dimensionless matrix, the neutrino mass matrix $\mathcal{M}$ in Eq.~\ref{eq:mass_matrix} can be block-diagonalized perturbatively.
At leading order, the $3\times3$ light-neutrino mass matrix is given by
the usual seesaw expression,
\begin{equation}
    m_\nu
    \simeq
    -m_D M_N^{-1}m_D^T = -\epsilon M_N \epsilon^T \,.
    \label{eq:seesaw_light_mass}
\end{equation}
The standard type-I seesaw limit is most transparent in the mass basis
of the sterile states. The symmetric matrix $M_N$ in Eq.~\ref{eq:sterile_mass_matrix} can be diagonalized
through a rotation $V_N$ as
\begin{equation}
    V_N^T M_N V_N
    = \operatorname{diag}(\tilde m_{N_1},\tilde m_{N_2}) \simeq  \operatorname{diag}(m_{N_1},m_{N_2}) \,,
\end{equation}
where the last equality holds at order zero in $\epsilon$.
Assuming for simplicity a real $M_N$, the masses of the two mostly sterile states are given by
\begin{equation}
    m_{N_{1,2}}
    \simeq
    \left|
    \sqrt{
        M^2+\frac{(\mu_1-\mu_2)^2}{4}
    }
    \mp
    \frac{\mu_1+\mu_2}{2}
    \right|\,.
    \label{eq:sterile_masses}
\end{equation}
Redefining the Dirac mass matrix as
$\widetilde m_D\equiv m_D V_N$, the light-neutrino mass matrix becomes
\begin{equation}
    (m_\nu)_{\alpha\beta}
    \simeq
    -\sum_{i=1}^{2}
    \frac{
        (\widetilde m_D)_{\alpha i}
        (\widetilde m_D)_{\beta i}
    }{
        m_{N_i}
    } \,.
    \label{eq:typeI_general}
\end{equation}

The next step (order $\epsilon$) in the perturbative diagonalisation
defines the active-sterile mixing matrix, according to
\begin{equation}
U_{\nu_\alpha N_i}
\simeq
\frac{(\widetilde m_D)_{\alpha i}}{m_{N_i}}\,.
\label{eq:seesaw_mixing}
\end{equation}
Substituting this relation into Eq.~\eqref{eq:typeI_general}, the
mass matrix of light neutrinos can be written as
\begin{equation}
(m_\nu)_{\alpha\beta}
\simeq
-\sum_{i=1}^{2}
m_{N_i}\,
U_{\nu_\alpha N_i}U_{\nu_\beta N_i} \,.
\label{eq:mnu_theta}
\end{equation}

The Casas--Ibarra parametrization provides a general solution for $U_{\nu_\alpha N_i}$ in terms of the light-neutrino mass matrix, the heavy-neutrino masses and a complex orthogonal matrix $R$~\cite{Casas:2001sr,Moffat:2017feq}.
Indeed, the Dirac mass matrix can be expressed as
\begin{equation}
\widetilde m_D
\simeq iU_\nu^*\sqrt{D_\nu}\,R\,\sqrt{D_N}\,,
\end{equation}
where $D_\nu=\operatorname{diag}(m_{\nu_1},m_{\nu_2},m_{\nu_3})$,
$D_N=\operatorname{diag}(m_{N_1},m_{N_2})$, $R$ is a
$3\times2$ complex orthogonal matrix satisfying
$R^TR=\mathds{1}_{2\times2}$, and $U_\nu$ is the PMNS matrix diagonalising $m_\nu$.

For two sterile neutrinos and normal ordering, the lightest
neutrino is massless, $m_{\nu_1}=0$, and $R$ can be
parametrized by a single complex angle $\omega$ as
\begin{equation}
R=
\begin{pmatrix}
0 & 0\\
\cos\omega & -\sin\omega\\
\sin\omega & \cos\omega
\end{pmatrix}.
\end{equation}
Using Eq.~\eqref{eq:seesaw_mixing}, the active--sterile
mixing can then be written as
\begin{equation}
\begin{aligned}
 U_{\nu_\alpha N_1}
 &\simeq\frac{i\left[
 \sqrt{m_{\nu_2}}(U_\nu)_{\alpha 2}^{*}\cos\omega
 +\sqrt{m_{\nu_3}}(U_\nu)_{\alpha 3}^{*}\sin\omega
 \right]}
 {\sqrt{m_{N_1}}}\,,\\
 U_{\nu_\alpha N_2}
 &\simeq\frac{i\left[
 -\sqrt{m_{\nu_2}}(U_\nu)_{\alpha 2}^{*}\sin\omega
 +\sqrt{m_{\nu_3}}(U_\nu)_{\alpha 3}^{*}\cos\omega
 \right]}
 {\sqrt{m_{N_2}}}\,.
\end{aligned}
\label{eq:CI_mixing}
\end{equation}

For normal ordering, the two non-zero light-neutrino
masses are $m_{\nu_3}=\sqrt{\Delta m_{31}^2}\simeq 50~\mathrm{meV}$ and $m_{\nu_2}=\sqrt{\Delta m_{21}^2}\simeq 8.6~\mathrm{meV}$,
using the best-fit values from NuFIT~\cite{Esteban:2024eli}.
Since $(U_\nu)_{e3}$ is suppressed by the small value of $\theta_{13}$, one finds
$|U_{\nu_e N_1}|\ll |U_{\nu_\mu N_1}|\approx |U_{\nu_\tau N_1}|$,
except for a small region of parameters where $|\sin\omega|\ll|\cos\omega|$. In the GeV mass region, we expect that setting $U_{\nu_e N_1}=0$ or allowing a small nonzero value has a negligible impact on our sensitivity results.
For this reason, for our study we adopt a simplified benchmark,
$U_{\nu_e N_1}=0$ and
$|U_{\nu_\mu N_1}|=|U_{\nu_\tau N_1}|$, corresponding to a
$0:1:1$ flavour structure, also considered e.g.~in
Ref.~\cite{Drewes:2022akb}.

To derive the  minimal seesaw band, we define a global $N_1$ mixing as
\begin{equation}
 U_{N_1}^2\equiv\sum_{\alpha=e,\mu,\tau}|U_{\nu_\alpha N_1}|^2
 =\frac{m_{\nu_2}|\cos\omega|^2+m_{\nu_3}|\sin\omega|^2}{m_{N_1}}\,,
 \label{eq:seesaw_total_mixing_N1}
\end{equation}
where the interference term vanishes by PMNS unitarity,
$\sum_\alpha(U_\nu)_{\alpha2}^*(U_\nu)_{\alpha3}=0$.
We restrict ourselves to real $\omega$: the active--sterile mixing can be exponentially enhanced by taking $\omega$ with a large imaginary part, but this amounts to a strong cancellation between the contributions of the two sterile neutrinos in Eq.~\eqref{eq:mnu_theta}, while here we stand by the generic seesaw
expectation.
Using $\cos^2\omega+\sin^2\omega=1$ and
$m_{\nu_2}<m_{\nu_3}$, we then obtain
\begin{equation}
 \frac{m_{\nu_2}}{m_{N_1}}\leq U_{N_1}^2
 \leq\frac{m_{\nu_3}}{m_{N_1}}\,.
 \label{eq:seesaw_total_band_N1}
\end{equation}
For an approximate $0:1:1$ flavour structure, one has
$U_{N_1}^2\simeq 2|U_{\nu_{\mu} N_1}|^2$, and hence
\begin{equation}
\frac{m_{\nu_2}}{2m_{N_1}}\lesssim
 |U_{\nu_{\mu} N_1}|^2\lesssim
 \frac{m_{\nu_3}}{2m_{N_1}}\,.
 \label{eq:seesaw_mu_band_N1}
\end{equation}
This interval defines the seesaw band
used in this work.

\section{Experiments and rates}
\label{app:widths_production}

In this appendix, we summarise the relevant experimental assumptions, the meson-production rates, and the heavy-neutrino decay rates, which determine the sensitivities presented in Sec.~\ref{sec:seesaw}.

\subsection{Input from the experiments}

We collect the experimental features of the four experiments considered in this work: CHARM, BEBC, T2K, and SHiP.

\paragraph{\textbf{CHARM:}}
The CHARM beam-dump experiment used $400$ GeV protons
with a total exposure of $2.4\times10^{18}$ POT.
The decay volume had dimensions
$3\times3\times35\,\mathrm{m}^3$, starting $480$ m
downstream of the target, with its center displaced
by $5$ m from the beam axis~\cite{Bergsma:1986is,Boiarska:2021charm}.
The search considered the leptonic decay channels
$N\to\nu e^+e^-$, $N\to\nu\mu^+\mu^-$, and
$N\to\nu e^\pm\mu^\mp$.
No candidate events were observed, 
thus excluding with $90\%$ C.L. the region of parameters with more than $2.3$ signal events,
assuming negligible background. Following Ref.~\cite{Boiarska:2021charm}, we adopt the channel-dependent reconstruction efficiencies
$\epsilon_{ee}=0.60$, $\epsilon_{e\mu}=0.65$, and
$\epsilon_{\mu\mu}=0.75$.

\paragraph{\textbf{BEBC:}}
The WA66 beam-dump experiment used $400$ GeV protons
with a total exposure of $2.72\times10^{18}$ POT.
The neon--hydrogen bubble chamber had a fiducial volume
of approximately $3.57\times2.52\times1.85\,\mathrm{m}^3$,
located $404$ m downstream of the beam dump
~\cite{CooperSarkar:1985nh,Barouki:2022bebc}.
The search considered the leptonic decay channels
$N\to\nu e^+e^-$, $N\to\nu e^\pm\mu^\mp$, and
$N\to\nu\mu^+\mu^-$, as well as the semileptonic
channels $N\to e^\pm\pi^\mp$ and
$N\to\mu^\pm\pi^\mp$. 
We follow the kinematic selections and adopt a $96\%$
identification efficiency from Ref.~\cite{Barouki:2022bebc}.
No candidates were observed in the leptonic or $e\pi$
channels, while one event was observed in the $\mu\pi$
channel, with an expected background of $0.6\pm0.2$ events.
We implement the corresponding $90\%$ C.L. upper limits of
$2.3$ and $3.5$ signal events, respectively.

\paragraph{\textbf{T2K/T2HK:}}
The T2K search used the gaseous TPCs of the ND280 near detector,
located $280$ m downstream of the target at an off-axis angle
of $2.5^\circ$, with a total gas volume of interest of
$6.3\,\mathrm{m}^3$~\cite{T2K:2019jwa}.
The analysed exposures were $12.34\times10^{20}$ and
$6.29\times10^{20}$ POT in neutrino and antineutrino mode,
respectively. The search considered HNLs produced in charged-kaon
decays and decaying into $\nu e^+e^-$, $\nu\mu^+\mu^-$,
$\nu e^\pm\mu^\mp$, $e^\pm\pi^\mp$, or $\mu^\pm\pi^\mp$.
The selection required two oppositely charged tracks forming a
vertex inside the TPC gas, with particle identification and
kinematic cuts described in Sec.~III~B of Ref.~\cite{T2K:2019jwa}.
We use the released mass- and channel-dependent selection
efficiencies. No candidates were observed in any signal category;
the residual background expectations are given in Table~II of
Ref.~\cite{T2K:2019jwa}. For the T2HK projection, we assume $2.7\times10^{22}$ POT,
retaining the same
ND280 fiducial volume and selection efficiencies, scaling signal
and background yields with exposure.

\paragraph{\textbf{SHiP:}}
We use the $400$ GeV SPS beam-dump benchmark with
$2\times10^{20}$ POT~\cite{SHiP:2018xqw,SHiP:2021nfo}.
The geometrical acceptance is approximated by a widening decay
volume between $60$ and $100$ m from the target, with entrance
and exit transverse dimensions $5\times10$ and
$10\times20\,\mathrm{m}^2$, respectively. 
For the selected charged-track modes we assume unit efficiency
and negligible background. 

\subsection{Argon TPCs at DUNE and SHiP}
\label{app:argon_tpcs}

Argon TPCs offer alternative possibilities for detecting
photons through conversion into charged pairs.
In particular, the Liquid-Argon Near Detector (ND-LAr)
of DUNE, foreseen for Phase I, provides
a dense medium in which photons have an attenuation
length of approximately $20$ cm, allowing efficient
photon conversion and reconstruction~\cite{Coloma:2023oxx,DUNE:2021tad}.
However, searches for $e^+e^-$ pairs face significant
backgrounds from neutrino interactions, particularly
neutral-current events producing $\pi^0\to\gamma\gamma$,
where only one photon converts in the detector and
mimics an $e^+e^-$ signal.
In contrast, $\mu^+\mu^-$ final states are expected
to have much smaller backgrounds, potentially background-free~\cite{Coloma:2023oxx}.

A more favorable environment for measuring $e^+e^-$
pairs could be provided by the proposed high-pressure
gaseous-argon detector (ND-GAr) for DUNE Phase II
~\cite{DUNE:2024wvj}.
Unlike the gaseous TPCs of T2K, ND-GAr is designed
to operate at approximately $10$ bar, increasing the
photon-conversion probability.
Although the photon attenuation length exceeds $10$ m,
conversion within the detector could provide a larger
signal contribution than the suppressed off-shell
decay $N_1\to\nu e^+e^-$.
Moreover, its lower density substantially reduces the
number of neutrino interactions in the detector~\cite{Coloma:2023oxx}. We will not include DUNE in our sensitivity curves, as projections for ND-LAr as robust as those for T2K would require dedicated studies of background rejection, while ND-GAr remains a proposal. Neither detector is expected to be competitive with SHiP for these searches.

The possibility of a liquid-argon detector at SHiP has also been discussed~\cite{Ferrillo:2023hhg}, which would make it possible to search directly for the dominant radiative decay $N_1\to\nu\gamma$ through photon conversion in the TPC. However, the proposed liquid-argon TPC would be substantially smaller than the SHiP decay volume, and neutrino backgrounds are also expected to be significant. Therefore, for dipole searches at SHiP, the suppressed off-shell channel $N_1\to\nu\ell^+\ell^-$ remains the more advantageous option, using the currently approved SHiP infrastructure.

\subsection{Mesons contributing to sterile-neutrino production}

\begin{figure*}[t]
    \centering
    \includegraphics[width=0.48\textwidth]
    {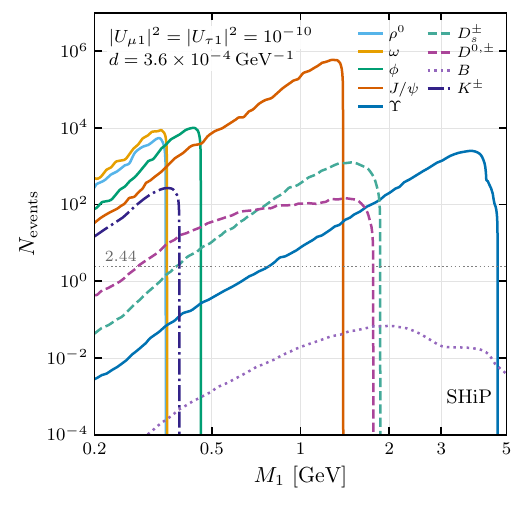}
    \hfill
    \includegraphics[width=0.48\textwidth]
    {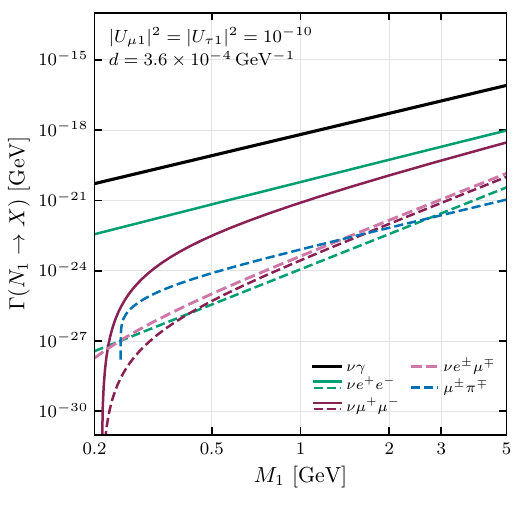}

    \caption{
    Left: Contributions from individual parent mesons to the
    expected number of $N_1$ decays at SHiP, summed over the visible decay channels shown in the right panel (all but $\nu\gamma$). The dotted
    horizontal line at $N_{\rm events}=2.44$ corresponds to a
    background-free $90\%$ C.L. sensitivity. Right: Main partial
    decay widths of $N_1$, induced by dipole+mixing (solid lines) or mixing only (dashed lines).
    For the $\nu \ell^+\ell^-$ channels, the two contributions are shown separately in the same colour.
    Both panels correspond to the benchmark
    $U_{e1}=0$,
    $|U_{\mu1}|^2=|U_{\tau1}|^2=10^{-10}$,
    $d=3.6\times10^{-4}\,\mathrm{GeV}^{-1}$, and
    $\delta=0.1$.
    }
    \label{fig:parent-events-and-decay-widths}
\end{figure*}

The left panel of Fig.~\ref{fig:parent-events-and-decay-widths}
illustrates the contributions from different parent mesons to the SHiP HNL event rate, 
for the benchmark parameters $U_{e1}=0$,
    $|U_{\mu1}|^2=|U_{\tau1}|^2=10^{-10}$,
    $d=3.6\times10^{-4}\,\mathrm{GeV}^{-1}$, and
    $\delta=0.1$.
Dipole production proceeds through $V^0\to N_1N_2$, with
$V^0=\rho^0,\omega,\phi,J/\psi,\Upsilon$, whenever
$m_{N_1}+m_{N_2}<m_V$~\cite{Barducci:2022gdv,Bertuzzo:2024eds}.
These rates scale as $|d|^2$ and are independent from the active--sterile mixing.
The mixing-induced contribution arises from leptonic and
semileptonic decays of kaons, charm and beauty mesons~\cite{Atre:2009rg,Bondarenko:2018ptm,SHiP:2018xqw}. We use \textsc{Pythia} 8.3~\cite{Bierlich:2022pfr} to simulate the meson production yields and kinematic distributions.

\subsection{Sterile-neutrino decay channels}

We take into account the standard weak decay channels, generated by the
active--sterile mixing, using the partial widths of
Refs.~\cite{Atre:2009rg,Coloma:2020lgy,Bondarenko:2018ptm},
together with the dipole-mediated processes discussed in
Sec.~\ref{sec:monophoton} and Ref.~\cite{Bertuzzo:2024eds}.

In the right panel of Fig.~\ref{fig:parent-events-and-decay-widths},
we have selected the main decay channels entering the results
presented in the main text. For the large dipole considered here,
the total $N_1$ width is dominated by $N_1\to\nu\gamma$.
However, as already discussed, a monophoton signal requires
photon conversion to be detected, and therefore the sensitivity
in an experiment like SHiP is expected to be negligible.
Its off-shell-photon counterpart,
$N_1 \to\nu\gamma^*\to\nu\ell^+\ell^-$, is suppressed by the
electromagnetic coupling and by three-body phase space, but
provides a detectable charged-particle pair in the SHiP decay
volume. 
Neglecting the additional weak-gauge-boson contributions to the same process (i.e.~taking the limit of vanishing Fermi constant), the
decay width reads
\begin{equation}\label{eq:Gamma_N1tonue+e-}
\begin{aligned}
    \Gamma(N_1 \to \nu_\beta \ell^+\ell^-) & = \frac{\alpha\,c_W^2\, |U_{\nu_\beta N_1}|^2
    |d|^2\,M_1^3}{384\pi^2} \\ & \times I\left(\sqrt{1-4 \frac{m_\ell^2}{M_1^2}}\right)\,,
\end{aligned}   
\end{equation}
where $\alpha$ is the fine-structure constant, $c_W$ the cosine of the weak angle, and the function $I(x)$ is given by
\begin{align}
    \begin{aligned}
        I(x) & = \left(x^6 - 3 x^4 +3x^2 + 15\right) \log\left(\frac{1+x}{1-x}\right)\\
        & \qquad -2 x \left(x^4 + 8 x^2 +15\right)\,.
    \end{aligned}
\end{align}
Notice that, in the $m_\ell \ll M_1$ limit, $\sqrt{1-4 m_\ell^2/M_1^2} \simeq 1$ and the decay width is enhanced by a large logarithm. This is due to a collinear divergence in the differential decay width, $d\Gamma/dq^2 \propto q^{-4}$ where $q^2$ is the virtuality of the photon, that appears in the limit $m_\ell \to 0$. The divergence appears because, in the massless limit, $q^2 = 2 E_{\ell^+} E_{\ell^-} (1-\cos\theta)$, where $E_{\ell^\pm}$ are the $\ell^\pm$ energies and $\theta$ the angle between their 3-momenta, so that $q^2\to 0$ when $\theta \to 0$.

In contrast, the flavour-off-diagonal decay
$N_1\to\nu e^\pm\mu^\mp$ and the charged-current semileptonic
decay $N_1\to\mu^\pm\pi^\mp$ cannot be induced by a photon exchange, therefore the observation of such final states would directly prove that the active--sterile mixing is non-vanishing, as emphasized in Sec.~\ref{sec:seesaw}.

The decay $N_1\to\nu\pi^0$ can also provide a significant
contribution to the signal, with the $\pi^0$ subsequently decaying
into two photons. However, we do not include this channel in
our sensitivity results, since achieving a background-free
search for this final state is more challenging. 

The above is not an exhaustive list of decay channels. Additional
modes, including $\mu^\pm\rho^\mp$, $\mu^\pm K^\mp$,
$\mu^\pm K^{*\mp}$, multihadron final states and,
above threshold, channels involving charmed mesons, are also
allowed~\cite{Atre:2009rg,Bondarenko:2018ptm}.
These channels can provide additional signals of
active--sterile mixing, although their more complicated
topologies, involving additional charged particles or photons
from secondary decays, make their reconstruction and background
rejection more challenging. Decays involving a primary $\tau$
additionally contain neutrinos from its subsequent decay. For the large dipole considered here, the contribution of these
additional channels to the total decay width is negligible,
since the latter is dominated by the dipole-mediated decay. Although these channels could provide additional
discriminators of active--sterile mixing, we conservatively
exclude them from our sensitivity estimates, which assume
perfect detection efficiency and zero background, due to the
additional challenges associated with their reconstruction.

\end{document}